\documentclass[12pt]{iopart}

\usepackage{supercite,amsfonts}
\usepackage[dvips]{graphicx}

\begin{document}

\title[Colloidal electrophoresis: Scaling]{Colloidal electrophoresis:
  Scaling analysis, Green--Kubo relation, and numerical results}

\author{B D\"unweg$^{1}$, V Lobaskin$^{1,2}$, K
  Seethalakshmy--Hariharan$^{1}$ and C Holm$^{1,3}$}

\address{$^1$ Max Planck Institute for Polymer Research, \\
  Ackermannweg 10, 55128 Mainz, Germany }

\address{$^2$ Physik-Department, Technische Universit\"at M\"unchen, \\
  James--Franck--Stra{\ss}e, 85747 Garching, Germany }

\address{$^3$ Frankfurt Institute for Advanced Studies (FIAS), J.~W.
  Goethe--Universit\"at, \\
  Ruth--Moufang--Stra{\ss}e 1, 60438 Frankfurt/Main, Germany }

\eads{\mailto{duenweg@mpip-mainz.mpg.de},
      \mailto{lobaskin@tum.de},
      \mailto{C.Holm@fias.uni-frankfurt.de}}

\begin{abstract}
  We consider electrophoresis of a single charged colloidal particle
  in a finite box with periodic boundary conditions, where added
  counterions and salt ions ensure charge neutrality. A systematic
  rescaling of the electrokinetic equations allows us to identify a
  minimum set of suitable dimensionless parameters, which, within this
  theoretical framework, determine the reduced electrophoretic
  mobility. It turns out that the salt--free case can, on the Mean
  Field level, be described in terms of just three parameters. A
  fourth parameter, which had previously been identified on the basis
  of straightforward dimensional analysis, can only be important
  beyond Mean Field. More complicated behavior is expected to arise
  when further ionic species are added. However, for a certain
  parameter regime, we can demonstrate that the salt-free case can be
  mapped onto a corresponding system containing additional salt. The
  Green--Kubo formula for the electrophoretic mobility is derived, and
  its usefulness demonstrated by simulation data. Finally, we report
  on finite--element solutions of the electrokinetic equations, using
  the commercial software package COMSOL.
\end{abstract}


\submitto{\JPCM}

\maketitle

\section{Introduction}
\label{sec:intro}

The behavior of charge--stabilized colloidal dispersions in external
electric fields is a classical topic of colloid physics \cite{Russel}.
Quantitative theoretical understanding is still incomplete today,
although substantial progress has been achieved over the decades
\cite{Smoluchowski,Hueckel,henry31a,wiersema66a,obrien78a,lozada99,%
  lozada01a,Carrique03}. The difficulty lies in the complicated
many--body nature of the problem, and hence only limiting cases are
well understood. Beyond the physics of the ``standard electrokinetic
model'' \cite{obrien78a}, which is essentially just a single--particle
Mean Field theory (see below), which nevertheless does describe a
quite broad range of phenomena, current research focuses mainly on
situations where this model is not applicable, or at least its
applicability is not obvious. These include cases where non--Mean
Field effects are important, i.~e. higher valency or non--negligible
size of the ions \cite{lozada99,lozada01a,tanaka01a}, or where the
single--colloid picture is expected to break down, due to overlapping
ionic clouds (or insufficient amount of screening by salt)
\cite{Carrique03}. This latter issue has also triggered detailed
experiments \cite{tom2,tom4,tom5,tom6,tom7} which measured the
electrophoretic mobility in the low--salt regime. Furthermore, the
problem has recently been studied by computer simulations
\cite{Yamamoto,apratim05,apratim07,NJP,JPCM,vladlett}. The
investigations of Ref. \cite{vladlett} were specifically targeted at
the low--salt limit. The purpose of the present paper is to provide
some more detailed theoretical and numerical background material which
had to be omitted in Ref. \cite{vladlett}. We will start in Sec.
\ref{sec:background} with a brief review of the simplest limiting
cases of electrophoresis, followed by a summary of the observations
made in Ref. \cite{vladlett}. The new material is then found in Secs.
\ref{sec:electrokin_eqs}--\ref{sec:COMSOL} (briefly outlined at the
end of Sec. \ref{sec:background}), while we conclude in Sec.
\ref{sec:Summary}.

\section{Background: Review of limiting cases, and
previous simulation results}
\label{sec:background}

\subsection{Single colloid without salt}

The simplest case of colloidal electrophoresis is obviously a single
charged sphere of radius $R$ and charge $Z e$ ($e$ denotes the
positive elementary charge, and we assume $Z > 0$), immersed in a
solvent of viscosity $\eta$ and dielectric constant $\epsilon$.
Assuming zero salt concentration, and zero colloidal volume fraction
$\Phi$, the drift velocity $\vec v$ which results as a response to an
external electric field $\vec E$ is just given by the Stokes formula,
\begin{equation}
6 \pi \eta R \vec v = Z e \vec E .
\end{equation}
This is the so--called H\"uckel limit \cite{Russel,Hueckel}. The
electrophoretic mobility $\mu$ of the colloidal sphere, defined
via
\begin{equation}
\vec v = \mu \vec E ,
\end{equation}
is hence given by
\begin{equation} \label{eq:mu1}
\mu = \frac{Z e}{6 \pi \eta R} .
\end{equation}
We now introduce the zeta potential as the electrostatic potential
at the colloid surface (with the understanding that it vanishes
infinitely far away from the colloid):
\begin{equation} \label{eq:zeta_simple}
\zeta = \frac{Z e}{4 \pi \epsilon R} .
\end{equation}
This allows us to re--write \eref{eq:mu1} as
\begin{equation} \label{eq:mu2}
\mu = \frac{2}{3} \frac{\epsilon}{\eta} \zeta .
\end{equation}
Based upon the thermal energy $k_B T$ as typical energy scale ($k_B$
denotes Boltzmann's constant and $T$ the absolute temperature), we can
introduce the dimensionless (reduced) zeta potential as
\begin{equation}
\zeta_{red} = \frac{e \zeta}{k_B T} .
\end{equation}
On the other hand, the thermal energy, combined with electrostatics,
provides a typical length scale, the Bjerrum length
\begin{equation}
l_B = \frac{e^2}{4 \pi \epsilon k_B T} ,
\end{equation}
which is nothing but the distance between two elementary charges such
that their electrostatic energy is just $k_B T$. This can be combined
with the Stokes formula to define a useful mobility scale for
electrokinetic phenomena:
\begin{equation} \label{eq:defmu0}
\mu_0 = \frac{e}{6 \pi \eta l_B} .
\end{equation}
Defining the reduced mobility as
\begin{equation}
\mu_{red} = \frac{\mu}{\mu_0} ,
\end{equation}
one sees that in the H\"uckel limit one simply has
\begin{equation} \label{eq:hueckel_mobility}
\mu_{red} = \zeta_{red} .
\end{equation}

\subsection{Zeta potential vs. reduced charge}

In a more general context, the electrostatic potential at the surface
of the colloid will of course no longer be given by
\eref{eq:zeta_simple}. It will rather be diminished, as a result of
the influence of the other charges. In order to clearly distinguish
between the concepts of charge and potential, we will call
\begin{equation} \label{eq:definetildeZ}
\tilde Z = \frac{Z e}{4 \pi \epsilon R} \frac{e}{k_B T}
         = Z \frac{l_B}{R}
\end{equation}
the reduced (re--parametrized) charge (regardless of the physical
situation), while the symbols $\zeta$ and $\zeta_{red}$ are reserved
for the actual value of the surface electrostatic potential and its
dimensionless counterpart. In the H\"uckel limit (and only in this
limit), $\zeta_{red}$ and $\tilde Z$ coincide.

\subsection{Screening}

An important aspect of electrophoresis is the screening of not only
electrostatic, but also hydrodynamic interactions. As soon as one
considers a system at a finite concentration, one has to take into
account that it must be charge--neutral, at least on sufficiently
large length scales: The charges (colloid charges plus ion charges)
that are contained in a sub--volume of linear dimension substantially
larger than the colloid--colloid correlation length must add up to
zero. The same is true (with arbitrary precision) in a computer
simulation if one considers the simulation box as a whole
(independently of the value of any correlation length).

Now, the basic mechanism of hydrodynamic screening is the fact that
the external electric field generates electric currents in \emph{both}
directions, which in turn generate hydrodynamic flows in both
directions. In leading order, however, these flows cancel, since the
total net force acting on the system (or sub--volume) is exactly zero.
As a result, the flow around a moving charged colloid will not decay
like $1/r$ (which would hold in the case of sedimentation, where the
system responds to a gravitational field and the net force does not
vanish), but much faster, $\sim 1/r^3$ \cite{Long}. The consequence
is, on the one hand, that finite size effects in computer simulations
are much less severe than in similar studies of sedimentation
\cite{tanaka01a}, and, on the other hand, that a single--particle
picture will apply whenever the electrostatic interactions are
sufficiently screened, as a result of high salt concentration. Indeed,
it is well--known that in the high--salt limit the screening of
electrostatics \cite{Russel} is governed by the Debye length $l_D =
\kappa^{-1}$, where the screening parameter $\kappa$ is proportional
to the square root of the salt concentration $c_s$:
\begin{equation} \label{eq:kappasalt}
\kappa^2 = 4 \pi l_B c_s .
\end{equation}
To be precise, \eref{eq:kappasalt} assumes monovalent salt ions, and
$c_s$ denotes the \emph{total} number of salt ions per unit volume
(such that the number of ion pairs per unit volume is given by
$c_s/2$). Now, under conditions where $l_D$ is substantially smaller
than the typical colloidal interparticle separation, it is clear that
most of the space between the colloids is charge--neutral.
Consequently, these regions are also force--free.  In other words, in
these regions there is no net flow, and all the hydrodynamic shear
gradients and viscous dissipation processes are confined to the Debye
layer as well. In this situation, one obviously can treat the problem
in terms of a single--particle picture. However, even the problem of a
single sphere surrounded by a charge cloud, with boundary condition of
vanishing electrostatic potential, and finite salt concentration, for
$r \to \infty$, can in general be solved only numerically. This is the
so--called ``standard electrokinetic model'' \cite{obrien78a}. The
reason for the mathematical difficulties is the non--linearity of the
underlying Poisson--Boltzmann equation, which determines the ionic
cloud structure.

\subsection{Smoluchowski limit}

A simple analytic solution is however possible in the limit of very
high salt concentration such that $l_D \ll R$. Here the geometry is
essentially planar, and one obtains the so--called Smoluchowski limit
\cite{Russel}:
\begin{equation}
\mu = \frac{\epsilon}{\eta} \zeta ;
\end{equation}
however, here the zeta potential is tiny, and in terms of the
reduced charge one has
\begin{equation} \label{eq:smoluchowski_mobility}
\mu_{red} = \frac{3}{2} \tilde Z \left( \kappa R \right)^{-1} .
\end{equation}
In the limit of infinitely strong screening ($\kappa \to \infty$), the
salt completely shields the electric field from the particle, and
correspondingly the mobility tends to zero. Of course, this is only a
mathematical limit, which can never be reached in practice, since at a
critical salt concentration the system of small ions will crystallize.
Beyond this critical concentration the liquid--state Smoluchowski
formula cannot work.

\subsection{Simulations of the low--salt case}

While the case of high salt concentration can thus be considered as
reasonably well understood (albeit in general only within the
framework of numerics), a completely different situation arises when
there is only little salt in the solution, or even none at all. In
this case the ionic clouds are mainly formed by the counterions, and
these will in general overlap. All the standard screening concepts,
which are based upon assuming a decay of the electrostatic potential
and of the charge density, on a length scale smaller than the
colloid--colloid separation, are no longer expected to
work. Nevertheless, namely due to the weak screening, some simplifying
assumptions can still be made for suspensions of strongly charged
colloids. As the colloids in this regime strongly repel each other,
they are usually well ordered so that their minimal separation amounts
to the mean interparticle distance $d \sim R \Phi^{-1/3}$. Thus, the
screening at $r < d$ will be exclusively due to counterions and the
phenomena that happen on this length scale will be governed by the
mean counterion concentration. These ideas proved to be useful for
describing static structure and colloidal interactions at low salt
\cite{rojas08}.

We have studied this case by computer simulations. In essence, our
method is Molecular Dynamics (MD) for the charged colloid, the
explicit (counter or salt) ions, and the solvent. However, for
computational efficiency the latter is replaced by a lattice
Boltzmann (LB) system which is coupled to the particles by a
Stokes friction coefficient.  This method, which has been designed
as an efficient and easy way of simulating systems with
hydrodynamic interactions, has been described in Refs.
\cite{Ahlrichs1,Ahlrichs2}, and is discussed in detail in a
forthcoming review article \cite{lbreview}. Langevin noise is
added to both the particles and the LB system to keep the
temperature constant. The colloid is modeled as a ``raspberry''
\cite{NJP,JPCM}, i.~e. a large central particle with a wrapping
consisting of a tether of small particles. The most important
results of this study have been communicated in Ref.
\cite{vladlett}, and can be summarized as follows:

(i) $\mu_{red}$ is a dimensionless quantity, and hence can only depend
on dimensionless parameters of the system. As a starting point, we
have made no further theoretical assumptions. In the salt--free case,
we can then identify four such parameters $p_1, \ldots, p_4$, which we
choose in such a way that two of them resemble most closely those
quantities which have proven useful in the ``salty'' case: these are
$p_1 = \kappa R$ and $p_2 = \tilde Z$ (cf. \eref{eq:hueckel_mobility}
and \eref{eq:smoluchowski_mobility}). In the present case, however,
$\kappa$ is not calculated from the salt concentration, but rather
from the counterion concentration:
\begin{equation} \label{eq:kappa_counterion}
\kappa^2 = 4 \pi l_B c ,
\end{equation}
with
\begin{equation} \label{eq:conc_counterion}
c = \frac{N Z}{V} = Z \frac{3}{4 \pi R^3} \Phi ,
\end{equation}
where $V$ is the system volume, and $N$ the number of simulated
colloids. Obviously, \eref{eq:kappa_counterion} and
\eref{eq:conc_counterion} imply the relation
\begin{equation} \label{eq:volfrac}
\left( \kappa R \right)^2 = 3 \tilde Z \Phi ;
\end{equation}
in other words, $\kappa R$ is nothing but a re--parametrized
volume fraction. It should be emphasized that due to assumed
strong charge asymmetry between the colloids and the counterions,
which both constitute the screening medium, the resulting charge
distribution is strongly inhomogeneous and the standard Debye
screening concept \emph{cannot} be implied. The remaining two
scaling variables are $p_3 = l_B / a$ and $p_4 = l_B / R$, where
$a$ is the counterion radius.

(ii) For a reasonable choice of parameters ($l_B / R$ not too large,
and $l_B / a$ of order unity, as is typical for ions in water) the
dependence on $p_3$ and $p_4$ can be ignored.

(iii) In terms of $\tilde Z$ and $\kappa R$, quite good agreement
can be achieved with experiments, provided $Z$ is replaced by an
effective charge, calculated from charge renormalization
\cite{rojas08,chargerenormalization,Belloni}.

(iv) Finite--size effects are weak, and hence one can obtain the data
for a certain finite volume fraction by just simulating a single
sphere in a suitably chosen finite box.

(v) The effect of added salt is similar to that of increased volume
fraction. It turns out that it is possible, within good approximation,
to just combine these two effects into one single parameter
\begin{equation} \label{eq:kappabar}
\bar \kappa^2 = 4 \pi l_B \left( c + c_s \right) ,
\end{equation}
which has a certain justification within a simplified linearized
Poisson--Boltzmann theory~\cite{beresford}.

The purpose of the present paper is to provide a theoretical
background for the observations reported in Ref. \cite{vladlett} and
derive some essential relations needed for the further analysis of the
electrophoresis at finite colloidal concentrations. In particular, we
describe our rescaling procedure in more detail. We feel that this
will become particularly transparent when done in terms of the
electrokinetic equations \cite{Russel}, which can be viewed as the
Mean Field description of the system we have simulated --- in contrast
to the simulation, the counterions are not considered as discrete
particles, but rather as concentration fields. Section
\ref{sec:electrokin_eqs} thus presents these equations, and outlines
the rescaling procedure. An important result of this analysis is that
the dependence on $p_4$ can indeed be ignored on the level of the
electrokinetic equations --- this parameter therefore describes
deviations from Mean Field behavior, if there are any. Section
\ref{sec:linresponse} discusses the problem of linear response, i.~e.
how to check that the non--equilibrium simulations employ a
sufficiently weak external field. We have solved this by comparing the
results with control calculations in strict thermal equilibrium, where
the mobility was calculated by Green--Kubo integration. As far as we
know, this formula has not yet been presented in the literature, and
we will derive it here. Finally, in Sec. \ref{sec:COMSOL} we present
some data which we have obtained by direct numerical solution of the
electrokinetic equations, using the commercial finite element package
COMSOL 3.3.

\section{Rescaling of the electrokinetic equations}
\label{sec:electrokin_eqs}

In the stationary state, the electrokinetic equations are given by
\begin{eqnarray}
\label{eq:incompressibility}
\nabla \cdot \vec v = 0 , \\
\label{eq:stokes}
- \nabla p + \eta \nabla^2 \vec v
- e (\nabla \Psi) \sum_i z_i c_i = 0 , \\
\label{eq:nernstplanck}
\nabla \cdot \left(
- D_i \nabla c_i - \frac{D_i}{k_B T} e z_i (\nabla \Psi) c_i
+ \vec v c_i \right) = 0 , \\
\label{eq:poisson}
\nabla^2 \Psi + \frac{1}{\epsilon} e \sum_i z_i c_i = 0 .
\end{eqnarray}
\Eref{eq:incompressibility} is the incompressibility condition for the
velocity field $\vec v$, while \eref{eq:stokes} is the Stokes equation
for zero Reynolds number flow, where the forces resulting from the
hydrostatic pressure $p$ and the viscous dissipation are balanced
against the electric force. Here, $\Psi$ denotes the electrostatic
potential, while $c_i$ is the concentration (number of particles per
unit volume) of the $i$th ionic species. We will adopt the convention
that $i = 0$ corresponds to the counterions, while $i \ge 1$ denotes
various types of salt ions. Each ion of species $i$ carries a charge
$z_i e$. Hence, the total charge density is given by $e \sum_i z_i
c_i$; this term appears also in the Poisson equation for the
electrostatic potential, \eref{eq:poisson}, where the boundary
conditions for $\Psi$ implicitly define the external driving field.
Finally, \eref{eq:nernstplanck} is the so--called Nernst--Planck
equation (convection--diffusion equation) which describes the mass
conservation of ionic species $i$. Here $D_i$ denotes the collective
diffusion coefficient of species $i$, while $D_i / (k_B T)$ is (via
the Einstein relation) the corresponding ionic mobility. The ionic
current consists of three contributions: the diffusion current, the
drift relative to the surrounding solvent, induced by the electric
force density $-e z_i c_i \nabla \Psi$, and finally the convective
current induced by the motion of the fluid.

We now introduce
\begin{equation}
M_i = \int d^3 \vec r \, c_i ,
\end{equation}
the number of ions of species $i$ in the solution, where the
integration extends over the finite volume $V$ of the system.
Obviously, the counterions just compensate the colloid charge, and
hence we have
\begin{equation} \label{eq:norm1}
- z_0 M_0 = Z ;
\end{equation}
note that $z_0 < 0$, and we consider only a single colloid in the
volume. Similarly, the charges of the salt ions compensate each
other, and hence
\begin{equation} \label{eq:norm2}
\sum_{i \ge 1} z_i M_i = 0 .
\end{equation}

In the case without external driving, we have $\vec v = 0$, and the
Stokes equation reduces to an equation which determines the pressure.
The Nernst--Planck equation, together with the Poisson equation, then
just becomes the Poisson--Boltzmann equation:
\begin{eqnarray}
\label{eq:pb1}
\nabla \ln c_i + z_i \nabla \tilde \Psi = 0 , \\
\label{eq:pb2}
\nabla^2 \tilde \Psi + 4 \pi l_B \sum_i z_i c_i = 0 ,
\end{eqnarray}
where we have introduced the abbreviation
\begin{equation}
\tilde \Psi = \frac{e \Psi}{k_B T} .
\end{equation}
In accordance with Ref. \cite{beresford} and \eref{eq:kappabar},
we introduce the parameter
\begin{equation} \label{eq:definekappa}
\bar{\kappa}^2 = 4 \pi l_B \frac{\sum_j z_j^2 M_j}{V} ,
\end{equation}
where however \emph{no} direct connection to a linearized
Poisson--Boltzmann equation is implied. We now use $\bar{\kappa}^{-1}$ as
our elementary unit of length and write
\begin{equation}
\nabla = \bar{\kappa} \tilde \nabla ,
\end{equation}
which transforms the Poisson equation into a fully non-dimensional
form:
\begin{equation}
\tilde \nabla^2 \tilde \Psi + \sum_i z_i \tilde c_i = 0 ,
\end{equation}
where non-dimensional concentrations $\tilde c_i$ are introduced via
\begin{equation}
c_i = \frac{\bar{\kappa}^2}{4 \pi l_B} \tilde c_i
    = \frac{\sum_j z_j^2 M_j}{V} \, \tilde c_i .
\end{equation}
In these scaled variables, the condition of mass conservation of
species $i$ is given by
\begin{equation}
\frac{1}{V} \int d^3 \vec r \, \tilde c_i =
\frac{M_i}{\sum_j z_j^2 M_j} = f_i
\end{equation}
(this equation defines the parameters $f_i$), where
\begin{equation}
\sum_i z_i^2 f_i = 1 .
\end{equation}
$\bar{\kappa}$ as a length unit also defines a dimensionless electric
field via
\begin{equation} \label{eq:dimensionlessfield}
\tilde{\vec E} = - \tilde \nabla \tilde \Psi
               = \frac{e}{\bar{\kappa} k_B T} \vec E ,
\end{equation}
and a dimensionless velocity $\tilde v$ by requiring that the relation
$v = \mu E$ transforms into $\tilde v = \mu_{red} \tilde E$:
\begin{equation}
\vec v = \frac{\bar{\kappa} k_B T}{6 \pi \eta l_B} \tilde{\vec v} .
\end{equation}
The diffusion coefficients $D_i$ can be mapped onto length scales
$a_i$ via a Stokes formula:
\begin{equation}
D_i = \frac{k_B T}{6 \pi \eta a_i} ,
\end{equation}
where $a_i$ is expected to be similar to the ion radius. Nevertheless,
it should be emphasized that the diffusion coefficient $D_i$ is a
\emph{collective} diffusion coefficient, not a tracer diffusion
coefficient. With these rescalings, the Nernst--Planck equation reads
\begin{equation}
\tilde \nabla \cdot \left(
- \frac{l_B}{a_i} \tilde \nabla \tilde c_i
- \frac{l_B}{a_i} z_i (\tilde \nabla \tilde \Psi) \tilde c_i
+ \tilde{\vec v} \tilde c_i \right) = 0 .
\end{equation}
Finally, we introduce a dimensionless pressure via
\begin{equation}
p = \frac{\bar{\kappa}^2 k_B T}{4 \pi l_B} \tilde p ,
\end{equation}
to re--write the Stokes equation in dimensionless form
\begin{eqnarray}
\tilde \nabla \cdot \tilde{\vec v} =  0 , \\
- \tilde \nabla \tilde p + \frac{2}{3} \tilde \nabla^2 \tilde{\vec v}
- (\tilde \nabla \tilde \Psi) \sum_i z_i \tilde c_i = 0 .
\end{eqnarray}
Let us collect the final set of non--dimensionalized equations:
\begin{eqnarray}
\tilde \nabla \cdot \tilde{\vec v} = 0 , \\
- \tilde \nabla \tilde p + \frac{2}{3} \tilde \nabla^2 \tilde{\vec v}
- (\tilde \nabla \tilde \Psi) \sum_i z_i \tilde c_i = 0 , \\
\tilde \nabla \cdot \left(
- \frac{l_B}{a_i} \tilde \nabla \tilde c_i
- \frac{l_B}{a_i} z_i (\tilde \nabla \tilde \Psi) \tilde c_i
+ \tilde{\vec v} \tilde c_i \right) = 0 , \\
\tilde \nabla^2 \tilde \Psi + \sum_i z_i \tilde c_i = 0 .
\end{eqnarray}

One sees that the only dimensionless parameters which remain in the
equations are the ratios $l_B / a_i$ and the charges $z_i$. Therefore,
in order to fully characterize the problem, one needs to specify three
parameters per ionic species ($l_B/a_i$, $z_i$, and $f_i$), plus the
parameters which pertain to the boundary conditions: The dimensionless
colloid radius $\bar{\kappa} R$, the dimensionless box size
$\bar{\kappa} L$ (note that we assume a cubic box with periodic
boundary conditions), and the non--dimensionalized charge density at the
colloid surface. For the latter, we note that in conventional units
the surface charge density is given by
\begin{equation}
\sigma = \frac{Z e}{4 \pi R^2} ,
\end{equation}
and that an electric field oriented perpendicular to the surface will
jump by a value $\sigma / \epsilon$. The jump in $\tilde E$ is
therefore given by $\tilde \sigma = e \sigma / (\bar{\kappa} \epsilon k_B
T)$, i.~e.
\begin{equation}
\tilde \sigma = \frac{\tilde Z}{\bar{\kappa} R} .
\end{equation}
Furthermore, \eref{eq:volfrac} is straightforwardly generalized in the
multi--ion case to
\begin{equation}
\Phi = \frac{- z_0 f_0}{3 \tilde Z} \left( \bar{\kappa} R \right)^2 ;
\end{equation}
this means that specification of $f_0$, $\tilde Z$, and $\bar{\kappa} R$
is enough to know $\bar{\kappa} L$.

We can thus summarize: In the case of zero salt and monovalent
counterions, the reduced mobility should be a function of just the
three parameters $p_1 = \bar{\kappa} R$, $p_2 = \tilde Z$, and $p_3 =
l_B / a$. This result should be contrasted with straightforward
dimensional analysis, which was the basis of the treatment in Ref.
\cite{vladlett}. Here one does not assume the validity of the
electrokinetic equations, i.~e. the assumption that the ionic cloud
can be treated as a continuum field is not made. Rather, one starts
from the observation that the problem is fully characterized by the
seven parameters $k_B T$, $\eta$, $L$, $Z$, $R$, $l_B$, and $a$. We
then replace $\eta$ by $\mu_0$ (see \eref{eq:defmu0}), $L$ by $\bar
\kappa$ (see \eref{eq:definekappa}), $Z$ by $\tilde Z$ (see
\eref{eq:definetildeZ}), and $a$ by $l_B / a$. This results in a new
but equivalent set of parameters $k_B T$, $\mu_0$, $\bar \kappa$,
$\tilde Z$, $R$, $l_B$, and $l_B / a$. Finally, we replace $l_B$ by
$l_B / R$ and then $R$ by $\bar \kappa R$ to find the parameter set
$k_B T$, $\mu_0$, $\bar \kappa$, $\tilde Z$, $\bar \kappa R$, $l_B /
R$ and $l_B / a$. We are thus left with seven parameters, of which
$k_B T$, $\mu_0$, and $\bar \kappa$ are needed to define the
fundamental units of energy, time, and length, respectively. Since
$\mu_{red}$ is a dimensionless quantity, it must be a function of the
remaining four dimensionless parameters, which are $p_1 = \bar \kappa
R$, $p_2 = \tilde Z$, $p_3 = l_B / a$, and $p_4 = l_B / R$. Since
$p_1$, $p_2$ and $p_3$ have also been identified on the basis of Mean
Field theory (i.~e.  the electrokinetic equations), we can only
conclude that any non--trivial dependence on $p_4$ must be the result
of deviations from Mean Field theory, i.~e. (most likely) ion
correlation effects. As a matter of fact, the successful comparison
between simulation and experimental data for $\mu_{red}$ that was done
in Ref.  \cite{vladlett} exclusively focused on the dependence on
$p_1$ and $p_2$. The justification for this procedure is that (i)
$p_3$ is of order unity both in simulation and experiment, and that
(ii) this implies a moderate strength of electrostatics. This means
that ion correlation effects are expected to be weak, which in turn
means a rather weak dependence on $p_4$, and adequacy of a description
in terms of the electrokinetic equations.

In the case of added salt, there are further parameters which enter
the problem; however, in the degenerate case, which was simulated in
Ref. \cite{vladlett} and where all ion types have the same properties
--- i.~e. all ions are monovalent, and have all the same mobility or
$l_B/a$ --- there are effectively only two ion types (the positive and
negative ones), and the only additional scaling variable which enters
is $f_0$, which specifies the fraction of counterions relative to the
salt ions. Apparently, $\mu_{red}$ is only weakly dependent on $f_0$
over a wide parameter range \cite{vladlett}.

In the case of finite salt concentration, we can consider the limit
$f_0 \to 0$, which implies $\Phi \to 0$ or $L \to \infty$. In this
case, the present formulation converges towards the situation studied
in the ``standard electrokinetic model'' \cite{obrien78a}. In the case
of zero salt, it is not possible to perform the limit $f_0 \to 0$
within our rescaled formulation, since then $\bar{\kappa}^{-1} \to \infty$,
and this is not suitable for a length unit. However, the physics of
just a free colloid is anyways trivial (see Sec. \ref{sec:intro}).

\begin{figure}
\begin{center}
\includegraphics[clip,width=10cm]{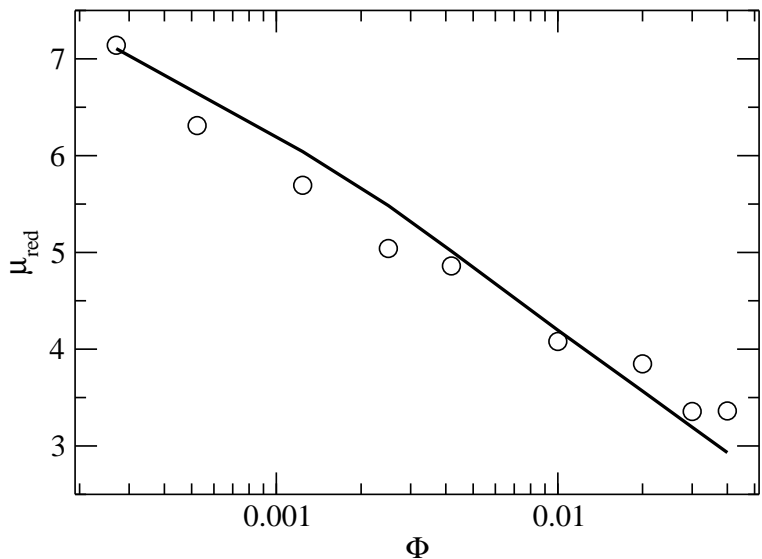}
\end{center}
\caption{Reduced electrophoretic mobility of the colloidal particle,
  as a function of the colloid volume fraction $\Phi$. The colloid
  charge is $Z=20$. The other parameters (in our simulational
  Lennard--Jones units) are: $\eta = 2.55$, $k_B T = 1$, $R = 3$
  (friction radius of the ``raspberry'', see text), $l_B = 1.3$, while
  the coupling of the small particles to the lattice Boltzmann fluid
  is characterized by a friction constant $\Gamma = 20$. For further
  details on the model, see Refs. \cite{NJP,JPCM}. The data
  points are simulation results. The solid curve is
  the H\"uckel formula prediction according to \eref{eq:psieff}.}
\label{fig:Huckel}
\end{figure}

Figure \ref{fig:Huckel} demonstrates again the general finding that
salt--free systems can, within reasonable approximation, be mapped
onto the corresponding ``salty'' system with the same $\tilde Z$ and
$\bar{\kappa} R$. For a dispersion with charge $Z = 20$, we compare the
simulation data for $\mu_{red}$, as a function of colloid volume
fraction $\Phi$, with the theoretical prediction that results from
this mapping. Since it turns out that in the simulated regime
of volume fractions $\bar{\kappa} R < 1$, it is reasonable to assume
that the H\"uckel formula \cite{Russel} holds:
\begin{equation} \label{eq:psieff}
\mu_{red} = \frac{\tilde{Z}_{\rm eff}}{1 + \bar{\kappa} R}=
\frac{\tilde{Z}_{\rm eff}}{1 + (3 \tilde{Z}_{\rm eff} \Phi)^{1/2}} .
\end{equation}
Here the factor $(1 + \bar{\kappa} R)^{-1}$ takes into account the
reduction of the surface potential, within the Debye--H\"uckel
approximation, while $Z_{\rm eff}$ was calculated via the charge
renormalization procedure by Alexander et
al. \cite{chargerenormalization}, based upon the Poisson-Boltzmann
cell model. A complication arises from the fact that our ``raspberry
model'' effectively defines \emph{two} radii: On the one hand, the
particles on the surface tether have a distance $R_1$ (here: $R_1 = 3$
in our Lennard--Jones units) from the colloid center. Since the
tethered particles are those that couple frictionally to the lattice
Boltzmann fluid, this is the hydrodynamic radius of the sphere. On the
other hand, the minimum distance between the ions and the colloid
center is one ion diameter larger, due to the repulsive interaction
between tether particles and ions. This defines $R_2 = 4$. It
therefore makes sense to calculate the volume fraction and the
$\bar{\kappa} R$ parameter on the basis of $R_2$, and to also use it
in the charge renormalization procedure. However, in the
transformation from $Z_{\rm eff}$ to $\tilde Z_{\rm eff} = Z_{\rm eff}
l_B / R$, we used $R_1$, in order to obtain the correct Stokes radius
in the limit $\bar{\kappa} \to 0$. This procedure yields good
agreement between simulation and theory, as seen from Figure
\ref{fig:Huckel}.  For the simulated $\Phi$ values, $Z_{\rm eff}$
varies between $16.2$ and $18.4$.

\section{Linear response}
\label{sec:linresponse}

In the present section, we will derive the Green--Kubo formula for the
electrophoretic mobility, which allows us to determine $\mu$ from pure
equilibrium simulations. To our knowledge, the formula has so far not
been presented explicitly in the literature; however, the derivation
is very straightforward within the framework of standard linear
response theory. We follow the approach of Ref. \cite{frenkelsmit},
which we find particularly transparent.

Starting point is the Hamiltonian
\begin{equation}
{\cal H} (\Gamma, t) =
{\cal H}_0 (\Gamma) + {\cal H}^\prime (\Gamma, t) =
{\cal H}_0 (\Gamma) - f(t) B (\Gamma) ,
\end{equation}
where ${\cal H}_0$ describes the unperturbed system, and $f(t)$ is a
weak external time--dependent field, which couples linearly to a
dynamical variable $B$. $\Gamma$ denotes the phase--space variable. We
are interested in the dynamic linear response of a variable $A$. The
time dependence of the mean value of $A$, $\overline{ A (t) }$, must,
for reasons of linearity and time translational invariance, have the
form
\begin{equation} \label{eq:linresp1}
\overline{ A (t) } = \left< A \right> +
\int_0^\infty d \tau \chi_{AB} (\tau) f(t - \tau) ,
\end{equation}
where $\left< \ldots \right>$ denotes the thermal average in the
absence of perturbations. The dynamic susceptibility $\chi_{AB}$ is
defined in such a way that it is zero for negative arguments; this
permits extension of the integration range in \eref{eq:linresp1} to
$(-\infty,+\infty)$.

For the special case that $f(t)$ is a constant $f_0$ for $-\infty < t
< 0$, and zero from then on, one has, for $t > 0$,
\begin{equation}
\overline{ A (t) } = \left< A \right> +
f_0 \int_t^\infty d \tau \chi_{AB} (\tau) ,
\end{equation}
or
\begin{equation} \label{eq:linresp2}
\frac{d}{dt} \overline{ A (t) } = - f_0 \chi_{AB} (t) .
\end{equation}

On the other hand, the statistical--mechanical expression
for $\overline{ A (t) }$ in such a ``switch--off experiment'' is
\begin{equation}
\overline{ A (t) } = \frac{ \int d\Gamma
\exp \left( - \beta {\cal H}_0 + \beta f_0 B \right) A(t) }
{\int d\Gamma
\exp \left( - \beta {\cal H}_0 + \beta f_0 B \right)} ,
\end{equation}
where $\beta = 1 / (k_B T)$, $A(t)$ denotes the time evolution of $A$
under the influence of ${\cal H}_0$ only, and the Boltzmann factor
describes the averaging over the initial conditions, which are
distributed according to the \emph{perturbed} Hamiltonian.
Linearizing this expression with respect to $f_0$ for weak
perturbations yields
\begin{equation}
\overline{ A (t) } = f_0 \beta
\left( \left< B(0) A(t) \right>
- \left< B \right> \left< A \right> \right)
\end{equation}
or
\begin{equation}
\frac{d}{dt} \overline{ A (t) } = f_0 \beta \left< B(0) \dot A(t) \right> .
\end{equation}
Comparing this with \eref{eq:linresp2} yields the correlation--function
expression for the dynamic susceptibility:
\begin{equation}
\chi_{AB} (t) = - \beta \left< B(0) \dot A (t) \right>
\end{equation}
for $t > 0$. Translational invariance with respect to time implies
\begin{equation}
0 = \frac{d}{dt} \left< B(t + \tau) A(t) \right> =
\left< \dot B(t + \tau) A(t) \right> +
\left< B(t + \tau) \dot A(t) \right> ,
\end{equation}
from which one concludes, by setting $\tau = -t$, the alternative (and
more useful) representation
\begin{equation} \label{eq:general_lin_resp_final}
\chi_{AB} (t) = \beta \left< \dot B(0) A (t) \right> .
\end{equation}
Considering the case that the external perturbation is completely
independent of time, and that $\overline A$ settles to a constant
value, one thus finds from \eref{eq:linresp1} and
\eref{eq:general_lin_resp_final}
\begin{equation}
\overline{A} = \left< A \right> + f_0 \beta \int_0^\infty dt \,
\left< \dot B(0) A (t) \right> .
\end{equation}
For the problem of electrophoresis, we consider a set of
particles $i$ with charges $z_i e$ at positions $\vec r_i$,
in an electric field $\vec E = E \hat{e}_x$. The perturbation
Hamiltonian is thus given by
\begin{equation}
{\cal H}^\prime = - E e \sum_i z_i x_i ,
\end{equation}
i.~e. $f_0 = E$ and $B = e \sum_i z_i x_i$. Denoting the velocity
of the $i$th particle in $x$ direction with $v_{ix}$, we thus
find
\begin{equation}
\dot B = e \sum_i z_i v_{ix} .
\end{equation}
On the other hand, we are interested in the velocity response
of one particular particle (say, $i = 0$), i.~e.
\begin{equation}
A = v_{0x} ,
\end{equation}
with $\left< A \right> = 0$. This yields directly the desired
Green--Kubo formula for the electro\-phoretic mobility
\begin{equation}
\mu = \frac{1}{3} \frac{e}{k_B T} \sum_i z_i \int_0^\infty dt \,
\left< \vec v_{i} (0) \cdot \vec v_{0} (t) \right> ,
\end{equation}
where we have averaged over the three spatial directions. It should be
noticed that the formula involves a mixed correlation between the test
particle and all charges, in contrast to the tracer diffusion
coefficient, which contains only the autocorrelation of the test
particle, and the electric conductivity, which involves the
autocorrelation of the collective current.

\begin{figure}
\begin{center}
\includegraphics[clip,width=10cm]{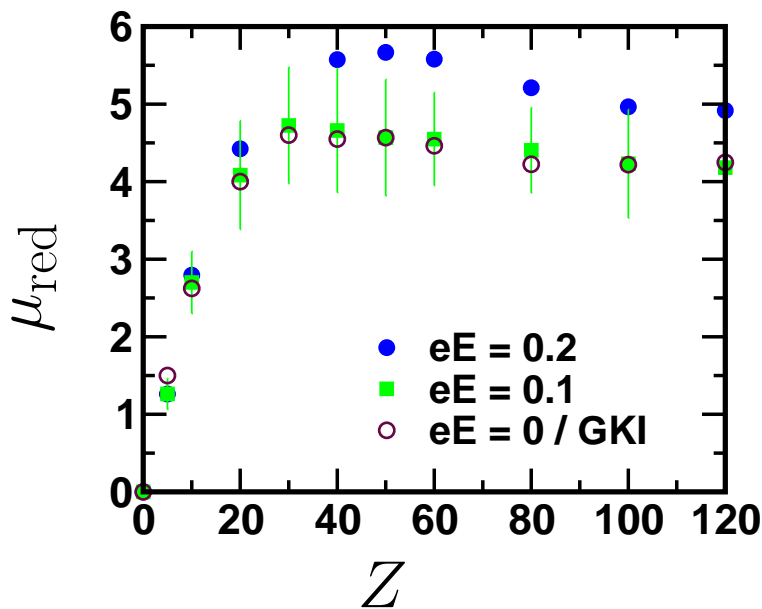}
\end{center}
\caption{Reduced electrophoretic mobility of the colloidal particle,
  as a function of its charge $Z$. No salt is added, and, apart from
  the central colloid, the system comprises $Z$ monovalent
  counterions. The linear box size is $L = 30$. The other simulation
  parameters are the same as those given in Figure
  \ref{fig:Huckel}. The mobility was here defined just as the ratio
  between drift velocity and electric field.  For strong driving field
  $E$, one observes nonlinear effects, while the results for weak
  driving agree favorably with the results of Green--Kubo integration
  (GKI). Note that the driving field is here given in the Lennard--Jones
  unit system of the simulation. According to \eref{eq:dimensionlessfield},
  constant $E$ does not imply constant $\tilde E$, since $\bar \kappa$
  varies with $Z$.}
\label{fig:GKI}
\end{figure}

As an example, we present Figure \ref{fig:GKI}, where the reduced
mobility for a salt--free system is plotted as a function of colloidal
charge.  Comparison with the Green--Kubo integral makes it possible to
check whether data obtained under non-equilibrium conditions are still
within the linear regime or not. One sees that the mobility first
increases with the charge (as one would expect from the physics of the
free colloid), but then saturates at a finite value, as a result of
condensation of more and more counterions. The nonlinear effects
observed for stronger electric fields are mainly due to charge--cloud
stripping \cite{JPCM}, which increases the effective charge and thus
the mobility.

\section{Finite element calculations}
\label{sec:COMSOL}

\begin{figure}
\begin{center}
\includegraphics[width=10cm]{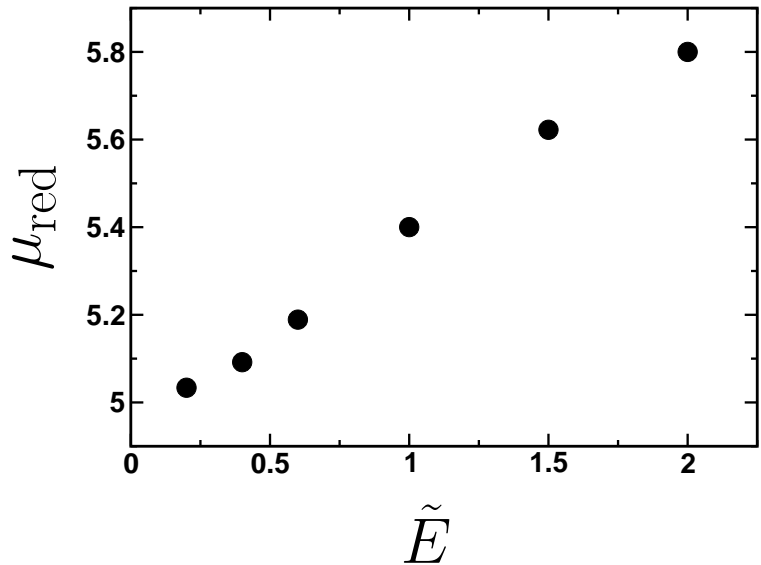}
\end{center}
\caption{Reduced mobility as a function of electric field $\tilde E$,
  for parameters chosen in accordance with those of Figure
  \ref{fig:GKI}, for charge $Z = 60$. Note that the electric field is
  given here in reduced units (see \eref{eq:dimensionlessfield}),
  i.~e. a value of $\tilde E = 1$ in the present plot corresponds to
  $e E \approx 0.2$ in the Lennard--Jones units of Figure
  \ref{fig:GKI}. }
\label{fig:comsol_mured}
\end{figure}

As a complementary approach to the hybrid MD~/~LB simulations, we have
also done some calculations where we solved the electrokinetic
equations directly, using a commercial finite--element software
package (COMSOL 3.3). For highly charged systems, where rather fine
grids are necessary, this does not work particularly well, since quite
generally the software tends to need excessive amounts of memory. We
used the same geometry as in the simulations, with the colloidal
sphere centered in the cubic box, but confined, for simplicity, the
computational domain to just the space outside the colloidal sphere.
This is not entirely correct, since, in reality, the electric field
also exists inside the sphere, where it takes a non--trivial
configuration, as a result of the external driving field oriented in
$x$ direction, the deformed charge cloud, and the cubic anisotropy.
However, if we assume that we can neglect the latter, and consider the
limit of infinitesimal driving, we get an electric field at the
colloid surface whose orientation is strictly radial, and whose value
is given by Gauss' law. This corresponds to the specification of
Neumann boundary conditions for the normal component of the electric
field. On the surface of the box, we specified Neumann boundary
conditions as well, where the normal component was set to zero in the
planes perpendicular to $y$ and $z$, while on the planes perpendicular
to $x$ it was set equal to the driving field. Concentration and flow
field were subjected to periodic boundary conditions. The pressure and
the electrostatic potential were set to zero at some arbitrary point
in the domain, in order to lift the degeneracy of shifting these
functions by an arbitrary amount. The Nernst--Planck equation was
augmented by a zero--flux condition at the colloid surface, and an
integral constraint in order to guarantee charge neutrality (such
integral constraints turn out to be computationally particularly
cumbersome). The flow velocity at the colloid surface was set to zero,
and the particle velocity was finally determined by transforming back
into the system's center--of--mass reference frame. Given the
inaccuracies of the boundary conditions, these results should not be
viewed as a stringent test of the validity of the Mean Field picture
for the simulated system. Nevertheless, the results agree reasonably:
Figure \ref{fig:comsol_mured} shows the reduced mobility as a function
of the external driving field, for a situation which corresponds to
the parameters of Figure \ref{fig:GKI} at charge $Z = 60$. In the
future, we hope to be able to calculate reduced mobilities in the
low--salt limit more easily by self--written software; efforts to
develop such a program are currently under way.

\section{Summary}
\label{sec:Summary}

In this paper we developed a theoretical basis for the scaling
analysis of the colloidal electrophoresis problem in the case of
finite colloidal concentrations. The rescaling procedure and
characterization of the dispersion in terms of effective dimensionless
parameters, i.~e. the reduced colloid charge, and the ratio of
screening length and size, allows one to map the numerical results
obtained for a single colloid onto experimental data for finite
colloidal volume fractions and no added salt. At least for a certain
parameter regime, we can also map the salt--free case onto a
corresponding system containing additional salt. Moreover, we
presented a numerically convenient method of measuring the colloidal
electrophoretic mobility based on the Green--Kubo analysis of the
equilibrium fluctuations of the charge motions. This allows for pure
equilibrium simulations and ensures that one always measures the
mobility and ion distributions in the linear response regime. Finally,
we gave an example of using a finite-element commercial software
package for solving the electrokinetic equations numerically, yielding
reasonable agreement with the simulations, and suggesting at least
consistency of the Mean Field picture with our simulations.

\ack{We thank T. Palberg for helpful discussions. This work was funded
  by the SFB TR 6 of the Deutsche Forschungsgemeinschaft.}

\section*{References}

\bibliographystyle{tryiop}
\bibliography{codef08}

\end{document}